\documentclass[runningheads]{llncs}
\usepackage{algorithm}
\usepackage{algorithmicx}
\usepackage[noend]{algpseudocode}
\usepackage{graphicx}
\usepackage{url}
\usepackage{array}
\usepackage{xcolor}
\newcolumntype{C}[1]{>{\centering\let\newline\\\arraybackslash\hspace{0pt}}m{#1}}
\usepackage{amssymb}

\newcommand*{\QEDA}{\hfill\ensuremath{\blacksquare}}%

\begin{document}
\title{VAPOR: a Value-Centric Blockchain that is Scale-out, Decentralized, and Flexible by Design}
\titlerunning{VAPOR}
%
\author{Zhijie Ren and Zekeriya Erkin}
%
\authorrunning{Z. Ren and Z. Erkin}
%
\institute{Department of Intelligent Systems \\
Delft University of Technology, The Netherlands \\
\email{\{z.ren, z.erkin\}@tudelft.nl}\\
}
\maketitle              
\begin{abstract}
Blockchains is a special type of distributed systems that operates in unsafe networks. In most blockchains, all nodes should reach consensus on all state transitions with Byzantine fault tolerant algorithms, which creates bottlenecks in performance. In this paper, we propose a new type of blockchains, namely Value-Centric Blockchains (VCBs), in which the states are specified as values (or more comprehensively, coins) with owners and the state transition records are then specified as proofs of the ownerships of individual values. We then formalize the ``rational'' assumptions that have been used in most blockchains. We further propose a VCB, VAPOR, that guarantees secure value transfers if all nodes are rational and keep the proofs of the values they owned, which is merely parts of the whole state transition record. As a result, we show that VAPOR enjoys significant benefits in throughput, decentralization, and flexibility without compromising security. 

\end{abstract}
\section{Introduction}\label{s:intro}
Blockchain technology, also referred as distributed ledger technology, considers a distributed system operating in a network with untrusted nodes. In blockchains, all nodes of the system apply the same rules to process consistent data, which mainly takes form of data blocks chained with unbreakable hash functions. We can categorize all existing blockchains into two categories by their data structures: one follows the idea of Bitcoin \cite{nakamoto} and we call Transaction-Centric Blockchains (TCBs), and the other follows from Ethereum \cite{ethereum} and the classical state machine replication model, we call Account-Centric Blockchains (ACBs). The former is commonly referred as ledgers, since all data are transactions, i.e., value transfer records. The concepts of account and balance are not explicitly emphasized. The latter, on the other hand, the states of nodes like their balances and other variables are defined and the state transition records, e.g., the transactions, are put to the back-end of the system. In either case, all nodes in the blockchain system should essentially always keep a consistent state regardless of whether the concept of state is explicitly emphasized. Then, in blockchains, nodes should only pre-agree with the initial state, i.e., the genesis block, and then be able to use a consistent rule to independently validate each input and then perform their state transitions.
As a result, both TCBs and ACBs require the complete state transition records to be acquired reliably and consistently by all nodes in the network, which causes a critical bottleneck in the performance of blockchain. In this paper, we use the term ``traditional blockchains'' to refer to all blockchains that all nodes need to acquire the whole state transition records.

A straightforward consequence of the bottleneck is the scalability issue which has been addressed in several other works \cite{croman,vukolic}. The throughput of blockchains does not grow with the number of nodes as the requirement of communication, computation, and storage grow at least proportionally to the number of nodes in the network. Hence, the throughput is limited to the capacity of the least capable node in the network and will not increase as the network grows.

Then, we also observe that centralization is an indirect consequence of the requirement for the whole state transition record.
As novel blockchain systems are pursuing high throughput in terms of transaction per second (TPS), the requirement for communication, computation, and storage becomes a threshold too high for normal users to participate. Then, the {\em participation threshold} is a crucial factor in evaluating the decentralization of the blockchain, since a blockchain with a high participation threshold will be consequentially unfriendly to normal users and more centralized, regardless of whether a fully decentralized consensus algorithm is used.

The third problem we address in traditional blockchains is inflexibility. As blockchains are decentralized by their nature, an upgrade or change to the system is much more difficult than centralized systems as inconsistency might happen if nodes follow different rules. Some examples of such inconsistency are ``forks'' like Bitcoin Cash/Bitcoin and Ethereum Classic/Ethereum, which cause the system to split and degrade in security.

In this paper, we address the problem of ``all nodes need to acquire and agree with all state transitions'' which essentially causes all above mentioned problems. To solve this problem, we propose a new type of blockchains called Value-Centric Blockchains (VCBs) that are equally secure as traditional blockchains but requires each node to only acquire partial state transitions. More precisely:
\begin{itemize}
\item We formalize the rationality of nodes in value transfer system, we call Rationality of Value Owner (RVO), which has already been explicitly or implicitly used in almost all blockchains without specification.
\item We propose a novel type of blockchains, called VCBs, which differ from traditional blockchains as the states are specified as the distribution for all values. A value can have an arbitrary amount and can be conceptually interpreted as a banknote. Then, all state transitions are sorted into proofs for the ownership of individual values. \footnote{Similar ideas can be found in many classical digital cash systems, i.e., Ecash \cite{ecash,ecash2}. The relationship and difference between VCBs and early digital cash systems will be discussed in Subsection~\ref{ss:vcp}.}
\item We propose a VCB called VAPOR in which nodes only needs to hold the proof of their own values. We further prove that it guarantees secure and fully decentralized value-transfer under the RVO assumption. Moreover, with examples, we show that VAPOR can be easily extended with extra functionalities like fast payment channels.
\item We show that VAPOR has significant advantages over traditional blockchains in throughput, decentralization, and flexibility.
\end{itemize}

This paper is organized as follows. In Section~\ref{s:ration}, we formally introduce the rationality of value owners in blockchains. Then, in Section~\ref{s:bc}, we introduce VCBs, their features, and the conditions required for a valid VCB. In Section~\ref{s:cvop}, we introduce a VCB, called VAPOR, and prove that it guarantees reliable value transfer. We show some examples of extension of VAPOR in Section~\ref{s:ecvop} and show the advantages of VAPOR over traditional blockchains in Section~\ref{s:adv}. At last, we compare our system to some related works in Section~\ref{s:rw} and conclude in Section~\ref{s:conc}.

\section{Rationality of Value Owner}\label{s:ration}
Blockchain technology is no stranger to the notion of rationality as it was introduced as one of the fundamentals of Bitcoin. However, the rational behaviors of nodes in blockchains, especially regarding the values they owned, are seldom formalized. 
A commonly utilized rationality assumption is that rational transaction issuers are motivated to prove to the receivers that the transactions are successful. It is mostly in the form of transaction fees, i.e., rational nodes would like to pay reasonable transaction fees so that their transactions could be added to the chains by the ``miners'', which is shown as the evidences that the transactions are successful. It has also been utilized in other forms, e.g., in the Tangle \cite{tangle}, rational nodes will do a POW and validate two previous transactions to make a transaction and in Omniledger \cite{omniledger}, rational nodes will take initiative in issuing their inter-shard transactions to all related shards and take effort in completing the transactions.

There is another type of rationality, the rationality of receiving values, which is mostly ignored in literature. In Bitcoin for instance, once a transaction is issued, a rational receiver should observe the chain for the transaction and a number of consecutive blocks to confirm the transaction. However, this is not emphasized since in most blockchains, the receiver do not need to validate extra information besides the blockchain itself. However, some off-chain solutions like Lightning Network (LN) and Plasma \cite{plasma,lightning} introduce new requirements for the rational receivers to validate some off-chain information to confirm a transaction.

Finally, we also specify a rationality, the rationality of holding values, which is usually considered trivial. In the basic Bitcoin system, it is simply holding the private key and keeping it secret. However, in current Bitcoin system, there are some special transactions called Pay-to-Script-Hash (P2SH) transactions, in which the values are locked by scripts that the value owners should be able to provide. 
Then, in LN, rational nodes also need to keep certain ``commitment transactions'' secretly. Moreover, they should actively monitor the chain to check if some specific transactions appear and take certain responses. Otherwise, their received transactions could be canceled.

In this paper, we formally introduce the Rationality of the Value Owners (RVO), which is the combination of all three rationalities mentioned above. These rules are in fact no stronger than the common rationality assumptions made in existing blockchains. We say that if a rational node follow the RVO rules, then he (we use the pronouns ``he'' for a node throughout this paper) would use his communication, computation, and the storage resources to perform the following:
\begin{itemize}
\item {\bf Rationality in Holding Value:} If he owns a value, he will make sure that he could prove the ownership.
\item {\bf Rationality in Sending Value:} If he sends a value, then he will take responsibility of proving to the receiver that 1), he owned this value; 2), the value is successfully transferred to the receiver.
\item {\bf Rationality in Receiving Value:} If he receives a value, then he will take responsibility of validating 1), the authenticity of that value; 2), the value transfer is successful. 
\end{itemize}

\section{Value Centric Blockchains}\label{s:bc}

The data structure of VCBs is similar to many ``off-chain'' schemes like \cite{rsk}. Each node individually puts its own transactions in off-chain transaction blocks and periodically sends an abstract of those blocks to a globally agreed main chain.
Then, the key elements in VCBs are values and their ownership. A value can be conceptually interpreted as a banknote with arbitrary denomination. Virtually, there exists a list of all values in the system, their amount, and their owners which updates with the system states. Moreover, for each ownership, there is a proof and an verification algorithm that could be used to determine the ownership, which consists of a subset of all transaction blocks. 
In this section, we introduce the basic concepts in VCBs: the main chain, the values, the verification algorithm, and the conditions required for a valid VCB, i.e., a valid VCB should be able to guarantee secure value transfers between nodes.

\subsection{Main Chain}

For a VCB, we define the {\em main chain} as a sequence of data blocks chained with unbreakable hash function, denoted by ${\cal B}=\{B_1, B_2, \ldots\}$. The main chain should have the following property, which is essentially achieved by all traditional blockchains.
\begin{property}[Consensus on the Main Chain]\label{pro:bft}
\begin{itemize}
\item {\bf Asynchronous Consistency:} In the situation where the message delay in the network is arbitrary, if an honest node agrees with a block $B_i$ as the $i$-th block of the chain, then another honest node will not agree with $B'_i \ne B_i$ as the $i$-th block of the chain.
\item {\bf Synchronous Liveness:} In the situation where the message delay in the network could be bounded by a constant $\tau$, if an honest node proposes a message $m$, then eventually an honest node will agree with a block $B$ containing $m$.
\end{itemize}
\end{property}
The main chain has two functions. First, it serves as a global clock. Throughout this paper, we use the term ``the system is at state $B_i$'' to represent a state that the system has just reached consensus on $B_i$. Second, it is used to reach consensus on data that needs global agreements, e.g., the initial value distribution, the verification algorithm, and digital signatures of the transaction blocks of nodes, which we will specify later.

\subsection{Value, Ownership, and Proof}\label{ss:vop}

We assume that there are $N$ nodes in the network, denoted by $1,2,\ldots,N$. We assume that there is a unique public key attached to each node and we can match the node and its public key when both are shown. In VCB, at each state of the system $B_i$, associated with a value $v_j, j=1,2,\ldots$, we have the amount of the value ${\cal Q}(B_1)=\{Q(v_1),Q(v_2),\ldots,\}$ and the owner of the value $O(v_j,B_i) \in \{NA,1,2 ,\ldots,N\}$. Here, $O(v_j,B_i)=NA$ suggests that this value is not owned by anyone at state $B_i$.
We define value distribution of state $B_i$ as ${\cal V}(B_i)=\{[v_j,O(v_j,B_i)]: \forall v_j \}$.
The initial value distribution and the amount of each value, i.e., ${\cal V}(B_1)$ and ${\cal Q}(B_1)$, are contained in the first block of the main chain $B_1$. Then, for a transaction, or more specifically a transfer of the value $v_j$ from owner $x$ to $y$, denoted by $tx_m(v_j, x\to y)$, we will have $O(v_j,B_i)=x$ and $O(v_j,B_{i+1})=y$ for a certain state $B_i$. Furthermore, we define a verification scheme, consists of an verification algorithm ${\tt GetOwner}(v_j,B_i, p)$ and proofs $P(v_j,B_i)$ for all $i,j$, that satisfies that 1), ${\tt GetOwner}(v_j,B_i,p)$ returns $O(v_j,B_i)$ if $p=P(v_j,B_i)$; 2), ${\tt GetOwner}(v_j,B_i,p)$ returns ``Fail'' if $p\ne P(v_j,B_i)$. The algorithm ${\tt GetOwner}(v_j,B_i, p)$ should also be agreed in $B_1$.


Now, we have all fundamental elements of VCBs: for a state $B_i$, there exists a set of values $v_j, \forall j$, their corresponding owners $O(v_j,B_i)$, their proofs of the ownership of the values $P(v_j,B_i)$, and an algorithm ${\tt GetOwner}(v_j,B_i,p)$ that could determine the owner of a value when the proof is given.

\subsubsection{Creating, Demolishing, Merging, and Dividing Values}\label{sss:cd}

The creation and demolition of values are crucial in many blockchains with Nakamoto-like consensus algorithms, since usually part of the incentives is given by creating new values. On the other hand, merging and dividing values are optional since the value exchange does not require the values to be divisible or mergeable, e.g., fiat currencies with banknotes and coins. Hence, we introduce how values could be created or demolished here, and the merging and dividing of values will be introduced in Section~\ref{ss:vd} as an additional functionality.

The creation and demolition of value should be agreed by all nodes, thus will be contained in the main chain. More precisely, to create a new value $v_j: [v_j, O(v_j, B_i)] \notin {\cal V}(B_i)$, a statement $[{\tt Add}:v_j,Q(v_j),O(v_j,B_{i+1})]$ should be in block $B_{i+1}$. Similarly, to demolish value $v_j$, we put a statement $[{\tt Delete}:v_j]$ in block $B_{i+1}$.

\subsection{Validity of VCB}
As far as we know, a rigorous definition of a valid value transfer system is still lacking, which remains a non-trivial and interesting topic for future research. In this work, we aim to propose a system that provides an equivalent value transfer functionality as other traditional blockchain systems, e.g., Bitcoin. Hence, we have the following definition for a valid VCB.
\begin{definition}[Valid VCB]\label{pro:vcb}
Firstly, we give the following properties.
\begin{itemize}
\item {\bf Ownership:} The owner of a value $v_j$ is able to validate the value and prove it to others, i.e., if $O(v_j,B_i)=x$, then node $x$ will eventually have $P(v_j,B_i)$. Moreover, the ownership can only be transferred by the owner.
\item {\bf Liquidity:} The owner of a value can transfer it to any other node within a certain period of time, i.e., if $O(v_j,B_i)=x$, then node $x$ can make $O(v_j,B_{i+k})=y$ for some $k, k\geq 1$.
\item {\bf Authenticity:} All values have at most one owner at each state, i.e., for all $v_j,B_i$, we have $O(v_j,B_i)\in \{NA,1,2,\ldots, N\}$.
\end{itemize}
A VCB is valid if and only if Ownership and Authenticity are guaranteed under asynchronous network settings and Liquidity is guaranteed in synchronous network settings.
\end{definition}


\subsection{RVO Rules in VCBs}

In a VCB, the RVO rules becomes:
\begin{itemize}
\item {\bf Rationality in Holding Value:} At a state $B_i$, if node $x$ is the owner of value $v_j$, he will always make sure that he has a proof $p$ such that ${\tt GetOwner}(v_j,B_{i}, p)=x$ unless he sends $v_j$ at $B_i$.
\item {\bf Rationality in Sending Value:} At a state $B_k$, for a value $v_j$ that $O(v_j,B_k)=x$, if node $x$ would like to send this value, he will take responsibility of providing to the receiver $y$: 1), the time of the transaction $B_i, i>k$; 2), a proof $p$ such that ${\tt GetOwner}(v_j,B_{i-1}, p)=x$ and; 3), a proof $p$ such that ${\tt GetOwner}(v_j,B_{i}, p')=y$.
\item {\bf Rationality in Receiving Value:} For node $y$ to receive this transaction, it will check 1), ${\tt GetOwner}(v_j,B_{i-1}, p)=x$; and 2), ${\tt GetOwner}(v_j,B_{i}, p')=y$.
\end{itemize}
\section{VAPOR}\label{s:cvop}

In this section, we propose a VCB, namely VAPOR, which stands for the five basic elements of our system, Value, Agreement, Proof, Ownership, and Rationality. As introduced in Section~\ref{s:bc}, a valid VCB should have the following.
\begin{itemize}
\item A main chain that guarantees Property~\ref{pro:bft}.
\item The owner and proof of value $O(v_j,B_i)$, and a valid authenticating scheme including $P(v_j,B_i)$ for all $i,j$ and a verification algorithm ${\tt GetOwner}(v_j,B_i,p)$ as described in Subsection~\ref{ss:vop}.
\end{itemize}
Now we describe these two parts in VAPOR. Then, we prove its validity and state its features.

\subsection{Main Chain and Its Consensus Algorithm}

There are two major types of algorithms that could achieve Property~\ref{pro:bft}: BFT algorithms and Nakamoto-like algorithms. The former includes \cite{castro,700bft,zyzzyva,miller} which explicitly requires the identity/public keys and the number of nodes to be predetermined and known by all nodes. The latter is inspired by Bitcoin and has been greatly developed in recent years. It contains a large number of algorithms such as Proof-of-Work based algorithms \cite{ng,byzcoin,hybrid}, Proof-of-Stake based algorithms \cite{snowwhite,algorand,ouroboros}, Directed Acyclic Graph based algorithms \cite{tangle,phantom,ghost}, etc. This type of algorithms do not require nodes to be predetermined. However, economical and game theoretical aspects have to be introduced to prevent Sybil attack as well as to encourage honest behaviors, and Property~\ref{pro:bft} is achieved with overwhelmingly high probability rather than absolute.


In VAPOR, any of the existing consensus algorithms that guarantee Property~\ref{pro:bft} (with a high probability) can be used for the main chain ${\cal B}=\{B_1, B_2, \ldots\}$. Then, VAPOR has the same requirements as the consensus algorithm and achieve the same level of security. For instance, if PBFT \cite{castro} is chosen, then VAPOR allows less than 1/3 of the predetermined nodes to be malicious. Then, if Bitcoin POW is chosen, then VAPOR tolerates less than 1/4 of the total mining power to be malicious \cite{majority} and the confirmation of the transactions is probabilistic.


\subsection{Proofs and the Verification Algorithm}

The main content of VAPOR is transactions. The proofs of the ownership of values are just different subsets of the whole transaction set. Here, we first introduce the data structure of the transactions, then introduce how the proof is chosen for each value.


\subsubsection{Transaction Blocks}\label{sss:ict}

In VAPOR, each node independently makes transaction blocks with the transacitons sent by itself. A transaction $tx_m(v_j,x \to y)$ is defined as $$tx_m(v_j,x \to y)=[v_j,y,sn],$$in which $sn$ is an internal serial number generated by node $x$ to identify his transactions. Since transactions are then put in blocks with index of $x$, $x$ is omitted in individual transactions. Note that here $m$ is a virtual global transaction identifier we used in this paper and it does not actually acknowledged by any node. Periodically, a node puts transactions in a transaction block $b$ and send an abstract, $$a(x)=[x,H(pk_x), Sig_x(x|H(pk_x)|MR(b))],$$to reach consensus on the main chain, where $H(pk_x)$ is the hash of the public key of $x$ and $Sig_x(H(pk_x)|MR(b))$ is a digital signature made with $H(pk_x)$ concatenated with the Merkle root of $b$ encrypted by the private key of $x$. In each round, at most one abstract from a node can be included in the main chain. If multiple different abstracts from the same node are received in the same round, then only one of them is considered valid. By the property of digital signature, the content of $b$ is immutable once the abstract $a(x)$ is confirmed on the main chain. Hence, we denote the abstract $a(x)$ contained in block $B_i$ by $a_i(x)$ and the block $b$ by $b_i(x)$ and call it a confirmed block. Then, as ${\cal B}$ is agreed by all nodes, blocks $b_i(x), \forall x$ will also form a chain that as immutable as ${\cal B}$. Then, we define ${\cal CB}=\{b_i(x), \forall i,x\}$.


\subsubsection{Transaction Fee for Abstracts}\label{ss:tf}

In our system, instead of individual transactions, the consensus is only reached on the abstracts. Then, for many consensus algorithms, a transaction fee should be provided to the block proposers, namely the miners, for them to include the abstract. The amount of the transaction fee should not be fixed so that a market can be created between the nodes and the miners. It can be achieved by introducing a new type of transactions in which the receiver is the miner, i.e., in a transaction block $b_i(x)$, node $x$ could create transactions in form of $tx_m(v_j,x \to [{\tt miner}])=[v_j,x,[{\tt miner}],sn]$, where $[{\tt miner}]$ is a variable that equals to the proposer of the block $B_i$.
A non-trivial problem for the transaction fee is that the sender of this transaction does not know the receiver in advance, which hinders him from sending the proof to the receiver. Hence, in the scope of this paper, the transaction fees are only feasible if the main chain uses BFT algorithms or algorithms that the block proposer is determined before the block, e.g., \cite{ng,algorand,ouroboros}. Then, the sender will give the proof of this transaction to the corresponding node so that the abstract would be included.


\subsubsection{Value Ownership and Proof}

Firstly, we define the ownership of values as the following.
\begin{definition}[Value Ownership]\label{def:vo}
\begin{itemize}
    \item The initial value ownership is agreed on the main chain, either by the initial value distribution in $B_1$ or value creation in $B_k, k\geq 1$.
    \item We assume that node $x$ started owning a value $v_j$ at $B_{i'}$. Then, he will transit the ownership of this value to node $y$ if he makes a transaction in a confirmed block $b_i(x)$ and has not make any transaction of this value in any confirmed blocks $b_k(x), k \in [i'+1,i-1]$.
    \item If there are more than one transaction of the same value in one transaction block, it is a clear sign of an attempt of double spending. Hence, we forbid this by stating that if a value is transacted more than once by its owner in a confirmed block, then the owner of that value is $NA$.
\end{itemize}
\end{definition}

Then, we define the proof $P(v_j,B_i)$ as a subset of ${\cal CB}$, which is essentially all confirmed transaction blocks that are considered in the second item of Definition~\ref{def:vo}, as well as all necessary public keys to verify them. The algorithm ${\tt Proof}(v_j,B_i,{\cal CB})$ can be used to get the proof $P(v_j, B_i)$, which is given in Appendix~\ref{a:proof}.

\subsubsection{Verification Algorithm}
Further, as defined in Subsection~\ref{ss:vop}, a verification algorithm in a VCB should be able to determine the ownership when the proof is given and output ``Fail'' if any input other than the correct proof is given. In Algorithm~\ref{alg:aa}, we propose ${\tt GetOwner}(v_j,B_i,p)$ that outputs $O(v_j,B_i)$ if $p=P(v_j,B_i)$ and outputs `Fail' for $p\ne P(v_j,B_i)$.
\begin{algorithm}
\caption{Verification Algorithm ${\tt GetOwner}(v_j,B_i,p)$}
\label{alg:aa}
\begin{algorithmic}
\State Get the block of initial distribution (creation) of value $v_j$ in the main chain: $B_{\tt index}$
\State Set ${\tt owner}$ according to the initial distribution from the main chain.
\State {\tt index}++;
\While{$a_{\tt index}({\tt owner})$ exists in $B_{\tt index}$}
    \If{$b_{\tt index}({\tt owner})$ or the public key of {\tt owner} does not exist in $p$}
        \Return Fail;
    \EndIf        
    \If{Merkle root and signature do not match}
        \Return Fail;
    \EndIf
    \State ${\tt count} \gets$ number of transactions of $v_j$ in $b_{\tt index}({\tt owner})$;
    \If{${\tt count}=0$}
        \State {\tt index}++;
    \ElsIf{${\tt count}=1$}
        \State {\tt index}++;
        \State ${\tt owner} \gets$ the receiver of the transaction of $v_j$;
    \Else
        \mbox{ }\Return Fail;
    \EndIf
    \If{${\tt index}>i$}
        \If{All data in $p$ are blocks and all blocks have been checked}
            \Return {\tt owner};
        \Else
            \mbox{ } \Return Fail;
        \EndIf
    \EndIf
\EndWhile
\end{algorithmic}
\end{algorithm}

The validity of ${\tt GetOwner}(v_j,B_i,p)$ as an verification algorithm could be easily shown. First, it uses the same method as the second item in Definition~\ref{pro:vcb} to check whether $p$ consists of the exact transaction blocks as $P(v_j,B_i)$ and any mismatch returns `Fail'. Then, since the algorithm use exactly the same rules as the definition of ownership to determined the owner, it returns $O(v_j,B_i)$ if $p = P(v_j,B_i)$.

\subsection{Validity of VAPOR}\label{ss:valvapor}

Here, we prove that VAPOR is a valid VCB under RVO rules and the consistency of the system is uncompromised even if RVO rules do not hold.
\begin{theorem}\label{th:valid}
In VAPOR, the properties of a valid VCB will hold in the following conditions.
\begin{table}[]
\centering
\begin{tabular}{C{2cm}||C{2cm}|C{2cm}|C{2cm}}

Properties & Ownership & Liquidity & Authenticity \\ \hline
Conditions & RVO rules & Synchrony &  --- \\ 
\end{tabular}
\end{table}
\end{theorem}

Due to space limitation, we only give an outline of the proof and provide the full proof in Appendix~\ref{a:prf}. The Ownership could be proved by induction: for each owner of the value, he is always able to receive the proof of the value from a rational previous owner. Moreover, only the owner can transfer the value since the transaction only happens when the block is confirmed. The Liquidity follows from the Synchronous Liveness property of the main chain. Then, the Authenticity follows from the Asynchronous Consistency of the main chain, which also guarantees the consistency of all confirmed transaction blocks. Then, Authenticity is proved as at each state, the values, owners, and proofs are based on the confirmed transactions blocks in a deterministic and one-to-one mapped fashion.

The holding condition of each property in Theorem~\ref{th:valid} provides a good insight on VAPOR and its differences from traditional blockchains. First, even if RVO rules do not hold, e.g., a sender refuses to send the proof to the receiver, it only causes a fail to prove the ownership of this exact value. The Liquidity and Authenticity of the system are not violated and other values are not corrupted. Second, the Ownership does not depend on synchrony. Hence, if a value is transferred and the network lose synchrony for Liquidity, the proof of the value could still be delivered to the receiver if the sender is rational.

\subsection{Features of VAPOR}\label{ss:imple}


The most distinctive feature of VAPOR is that each node only needs to acquire and keep the proofs of the values that it owns, i.e., at a state $B_i$, node $x$ only needs to have $P(v_j,B_i), \forall O(v_j,B_i)=x$.
To efficiently record the proofs, we propose the following implementation:
\begin{itemize}
    \item The main chain is stored and updated according to the consensus algorithm.
    \item A node keeps a transaction block database of for all confirmed transaction blocks that he has.
    \item A node keeps a value ownership table that updates with the main chain and keeps track of the values, their owners, and the proofs that he knows, which includes his own values. The proofs are simply pointers to the transaction block database.
\end{itemize}
Comparing to TCBs and ACBs, a transaction of multiple values need to be recorded as multiple transactions in VAPOR. However, for all these transactions plus all transactions included in the same transaction block, only one signature is required in VAPOR, which is in fact more efficient in storage.
The communication is also efficient as transaction blocks are acquired directly from the sender of the value with point-to-point communication and guaranteed security under the RVO rules. Moreover, the receivers could inform the sender about the transaction blocks that it already has to avoid overhead.
Then, as a trade-off between storage and communication, a node can choose to not delete the proofs of the already spent values. This means that they do not need to re-acquire some transaction blocks for future received values. 

\section{Extending VAPOR by Modifying the Verification Algorithm}\label{s:ecvop}

In Section~\ref{s:cvop}, we introduced how transactions could be verified with the verification algorithm ${\tt GetOwner}$ with the proof $P(v_j,B_i)$. In this section, we show the flexibility of this framework by providing examples of extended functionalities. More precisely, we will show that the functionalities of value division, fast off-chain transactions, and value-related smart contracts can be easily achieved by simple modifications to the verification algorithms.

\subsection{Value Division}\label{ss:vd}

The functionality of value division can be achieved with a new type of transactions called value division that has the form: $$[{\tt Divide}: v_{\tt source} \to (v_{{\tt source},1}, Q(v_{{\tt source},1})), \ldots, (v_{{\tt source},n}, Q(v_{{\tt source},n})).$$
The index ${\tt source}$ forms a chain that can be traced back to the origin. Then, to validate a value divided from another value, we simply call ${\tt GetOwner}$ to check the owner of each value on the chain recursively from the origin. This new type of transactions can either be added by making modifications to {\tt GetOwner} or defining another algorithm {\tt GetOwnerDV} on the main chain that recursively calls {\tt GetOwner}. We describe {\tt GetOwnerDV} in Appendix~\ref{a:godv}.

\subsection{Fast Off-chain Payment}\label{ss:fop}
In VAPOR, the confirmation of the transaction is dependent on the main chain, thus it essentially has the same latency as traditional blockchains. However, a fast off-chain payment solution like LN or Plasma \cite{plasma,lightning} can also be deployed in VAPOR. 
Briefly speaking, an off-chain payment scheme works as follows. Firstly, some value is locked on the main chain as the deposit for the ``fast payment channel'' to a particular receiver. Then, transactions can be made to that receiver without confirmations on the main chain. The safety of the transactions are guaranteed by a mechanism for the receiver to take all deposit when the sender tries to cancel a transaction. However, this mechanism requires synchrony between the receiver and the main chain. Then, there is a mechanism allowing the sender to safely shut the off-chain payment channel at any time.

In VAPOR, similar ideas can be implemented under the same synchrony assumption. A node can independently lock its values for a receiver and then makes off-chain transactions by signing them and sending signed transactions to the receivers as proofs. Then, the verification scheme should be modified to be able to verify these proofs. The detail of this scheme will be given in Appendix~\ref{a:offchain}.


\subsection{Smart Contracts}

In the previous subsections, it is revealed that additional functionalities can be easily achieved by changing the rules for verification, which is merely a modification to ${\tt GetOwner}$, or agreeing on new verification algorithms on the main chain. 
In fact, as long as values are transferred and there are interested parties following RVO rules, smart contracts can be written in VAPOR as new verification algorithms with one principle: only data that is against the value owners' interest is required to be put on the main chain and other data can be safely moved off-chain to the corresponding value owners.
We give an example of such smart contracts, a betting game, in Appendix~\ref{a:betting}.

\section{Advantages of VAPOR}\label{s:adv}
It has been shown that in VAPOR, nodes do not necessarily need to record the whole transaction set to allow secure value transfer. This fundamental difference from traditional blockchains leads to the advantageous in throughput, decentralization, and flexibility.

\subsection{Throughput}\label{ss:tp}
The most straightforward advantage of VAPOR is the throughput because nodes only need to acquire the proofs of their own values instead of the whole transaction set, as stated in Subsection~\ref{ss:imple}. However, this improvement is not trivial to quantify as it depends heavily on the networks and the transaction patterns. Here, we theoretically analyze the throughput in terms of the transaction cost $C$, defined as a combination of the expected bandwidth, computation, and storage resources required to communicate, validate, and store a transaction in the whole network.

Unlike traditional blockchains, the cost of an individual transaction in VAPOR is determined by the proof size, which is situational. Hence, we calculate $C$ by looking at the expected transaction blocks in a round that a node eventually needs to acquire, which we denote by $b$. Then, we have $C=O(b)$ since a transaction will be eventually acquired by $b$ nodes on average. Let us consider a transaction block $b_i(x)$. It will eventually be acquired by node $y$ if node $x$ holds a value at state $B_i$ and at a state $B_j, j>i$ node $y$ receives that value. In other words, for the set of values ${\cal V}_i(x)$ holding by node $x$ at state $B_i$, if all other nodes will receive a value from ${\cal V}_i(x)$ sometime in the future, then VAPOR have no throughput gain over traditional blockchains. In all other cases, as long as there exists some nodes that will never acquire any value in ${\cal V}_i(x)$, then we have $b<N$ and VAPOR has a throughput benefit.

In \cite{ren}, a concept of spontaneous sharding is proposed, which roughly works as the following. When performing a transaction, a rational node will choose the value with the least transaction blocks to transmit among all values that he has. In other words, they tends to use the values for which the most part of the proof is already known and validated by the receiver, e.g., the value that once owned by the receiver. As a result, some values will only cycle in a part of the network, namely a shard, instead of the whole network. Then, a node holding $g$ values is equivalent to participating in $g$ shards and $b$ will then equal to the expected size of the union of these shards. Then, it is shown in \cite{ren} that in many scenarios, we have $C=O(b)=o(N)$, i.e., the throughput will scale out.
Note that any group of frequent transacting nodes can decide to perform this optimization at any time to gain the throughput benefit, regardless of the rest of the network. Hence, since spontaneous sharding gives direct benefit to individuals even if other nodes refuse to cooperate, the ``the tragedy of the commons'' \cite{hardin} problem will not occur. We refer the readers to \cite[Remark 2]{ren} for more discussion.

\subsection{Decentralization}
In Section~\ref{s:intro}, we address the centralization problem due to the high participation threshold.
In VAPOR, this problem is significantly mitigated due to the value centric principle: nodes only transmit and store the data needed for validation of their own values, which is mostly not the whole transaction set. For example, in traditional blockchains, for nodes who only own a few coins in a blockchain, they still have to acquire and validate the whole chain to validate their own values and make transactions. In VAPOR, their cost of validating their own values and making transactions is $O(1)$. 

\subsection{Flexibility}\label{ss:flex}

As shown in Section~\ref{s:ecvop}, VAPOR enjoys benefits of easy modification, extension, and upgrading by simply agreeing on new verification algorithms on the main chain. However, this can be pushed one step further by allowing nodes to individually choose the algorithms that they like to use.
Then, hard forks like Bitcoin/Bitcoin Cash or Ethereum/Ethereum Classic can be avoided. Instead, the forks will be ``hidden'' as some values might not be validated by some users as they disagree with a certain rules. However, they could still agree with the main chain and contribute to the security of the entire system. We consider this as an advantage of flexibility, as nodes are more freely to agree/disagree with each other, without destroying the consistency of the whole system as long as they have the basic agreement.

\section{Related Works}\label{s:rw}

This work is mainly inspired and developed from \cite{ren}. However, it does has similarities to other studies if we view VAPOR in different perspectives. We explain the similarities and relations of this work and other works in this section.

\subsection{Value Centric Principle}\label{ss:vcp}
The origin of describing value transfer systems by values (or alternatively called coins, notes, bills) can be dated back to some pioneering digital cash works like \cite{ecash,ecash2,uec}. However, in these schemes, the notions of value and transaction are interchangeable as a central authority is required to validate each transaction. Hence, Bitcoin, as well as most of its successors known as alt-coins, use TCBs that focus on the validity of individual transactions rather than the value. The main difference from TCBs and VCBs can be clarified using the example of the Simple-Payment-Verification (SPV) nodes in Bitcoin. SPV nodes could verify whether all related transactions of a value are validated by the miners and are on-chain, but they could not validate the authenticity of this value, i.e., could not detect double-spending.

Chainspace \cite{chainspace} is a blockchain with sharding that uses a similar value-centric idea for inter-shard transactions, i.e., each transaction should include a ``Trace'' pointing back to the source of the value, so that the validators from the value-receiving shard only need to check the shards of the sources to prevent double spending. However, it has more redundancy as the value-centric idea is used in a shard level instead of the node level, and thus has less throughput improvement comparing to VAPOR.


\subsection{Off-Chain and DAG Techniques}
In the perspective of data structure, VAPOR has its similarities to many off-chain systems like RSK \cite{rsk} as data is stored off-chain and a main chain is used for the hash of the data. However, most off-chain systems compromise in decentralization as some trusted nodes are required to validate the contents of the off-chain data. Also, comparing to the off-chain payment schemes like LN and Plasma \cite{plasma,lightning}, VAPOR essentially moves all proofs for values off-chain. As a result, it is no longer necessary to use deposits to enforcing the consistency of the off-chain and on-chain values. Then, it is also similar to Hashgraph \cite{swirld} in the sense that node individually create their own transactions. However, in Hashgraph, all nodes eventually need the whole transaction set. 

\subsection{Sharding}
Recently, many sharding schemes have been proposed to divide the network into small shards. Then, the transactions in a shard do not need to be communicated outside the shard. However, a key problem is that the double spending prevention of inter-shard transactions relies on the security of shards instead of the whole network, which is a degradation in the security. Shards can be either determined artificially by the network topology \cite{ethshard} or at random \cite{omniledger,elastico}, or determined based on applications or users \cite{rchain,chainspace}, to reduce the number of inter-shard transactions as well as the probability of malicious shards. However, our system guarantees no degradation on security since essentially, the shards are spontaneously formed by the value transfer patterns. In other words, all shards are secure for their own intra-shard transactions and there will be no inter-shard transactions. 

\subsection{Performance Comparison}

It is difficult to make fair throughput comparison between VAPOR and other systems using a uniform standard, e.g., transaction per second (TPS), as schemes have different security assumptions and the throughput also depends on the network settings. Therefore, we use a theoretical approach to analyze and compare the throughput and security of VAPOR with a typical system of each kind, i.e., LN for off-chain schemes, PHANTOM for DAG, and Omniledger for sharding schemes.
We consider the transaction cost $C$ (defined in Subsection~\ref{ss:tp}) and the security $S$ of a transaction, which is defined as the amount of compromised nodes (corresponding resources for POW or POS) required to perform a double-spending attack.
We present the results in Table~\ref{tb:perf}.
\begin{table}
\centering
\begin{tabular}{C{2cm}||C{2cm}|C{2cm}|C{2cm}|C{2cm}}
Schemes & VAPOR & LN & PHANTOM & Omniledger \\ \hline
$C$ & $O(b)$ & $O(1)$ & $O(N)$ & $O(d)$ \\ \hline
$S$ & $O(N)$ & $O(1)$ or $O(N)$ & $O(N)$ & $o(N)$ \\ 
\end{tabular}
\caption{The cost and security of a transaction in VAPOR, LN, PHANTOM, and Omniledger for the whole network. Here $b$ is the average transaction blocks of each state acquired by a node and $d$ is size of the shard.}\label{tb:perf}
\end{table}

The cost and security of VAPOR are given in Subsection~\ref{ss:tp} and Subsection~\ref{ss:valvapor}, respectively.
For LN, note that this transaction is different from classical notion of transactions as it relies on a deposit and the value would be locked until the channel is shut down. The security relies on the synchrony between the receiver and the system (explained in Appendix~\ref{a:offchain}), thus would be compromised if either one is compromised. PHANTOM uses a block DAG structure to remove the dependency of security on the throughput of a chain-structure blockchain. However, all nodes still need to eventually acquire all transactions and the system will not scale out. Omniledger reduces the cost to $O(d)$ where $d$ is the shard size and promises a throughput benefit that is proportional to $N/d$. However, as Omniledger yields a random approach to keep the malicious nodes within each shard to be below 1/3, the security of the system becomes a non-trivial function of $d$ and $N$, which is dominated by $N$ but not explicitly stated in \cite{omniledger}.

\section{Conclusion}\label{s:conc}


In this paper, we address and formalize the fundamentals of a value-transfer system and the rationality assumptions. The highlight of this work is that we clarify the redundancy in traditional blockchains for value-transfer and how this redundancy can be removed by using the rationality assumptions and VCBs. We hope that this work would set a theoretical framework for future blockchain designs and inspire many theoretical studies on other basic concepts in blockchains, e.g., the rational assumptions in non-value-transfer blockchains.

%
%

\bibliographystyle{splncs04}
\bibliography{Implicit_Consensus}
\appendix

\section{Algorithm ${\tt Proof}(v_j,B_i,{\cal CB})$}\label{a:proof}
We define the proof of the ownership $P(v_j,B_i)$ as a subset of ${\cal CB}$ that output by an algorithm ${\tt Proof}(v_j,B_i,{\cal CB})$ shown in Algorithm~\ref{alg:proof}.
\begin{algorithm}
\caption{${\tt Proof}(v_j,B_i,{\cal CB})$}
\label{alg:proof}
\begin{algorithmic}
\State Get the block of initial distribution (creation) of value $v_j$ in the main chain: $B_{\tt index}$
\State Set ${\tt owner}$ according to the initial distribution from the main chain.
\State {\tt index}++
\State {\tt Proof}=\{\}
\While{$a_{\tt index}({\tt owner})$ exists in $B_{\tt index}$}
    \If{Merkle root and signature do not match}
        \Return {\tt Proof}
    \EndIf
    \State Add $b_{\tt index}({\tt owner})$ and the public key of {\tt owner} to {\tt Proof}
    \State ${\tt count} \gets$ number of transactions of $v_j$ in $b_{\tt index}({\tt owner})$
    \If{${\tt count}=0$}
        \State {\tt index}++
    \ElsIf{${\tt count}=1$}
        \State {\tt index}++
        \State ${\tt owner} \gets$ the receiver of the transaction of $v_j$.
    \Else
        \mbox{ }\Return {\tt Proof}
    \EndIf
    \If{${\tt index}>i$}
        \Return {\tt Proof}
    \EndIf
\EndWhile
\end{algorithmic}
\end{algorithm}

\section{Proof for Theorem~\ref{th:valid}}\label{a:prf}
\begin{proof}
Firstly, we prove Ownership by induction. It is clear that the first owner of any value $v_j$ will have the proof of this value, which are basically all of his public key and his own confirmed transaction blocks until the block before the one that spends it. Then, assume that the $t$-th owner of $v_j$, denoted by $o_t$, has the proof $P(v_j,B_k)$ proving the ownership $O(v_j, B_k)=o_t$ at state $B_k$. Then, assume that the $t+1$-th owner, $o_{t+1}$ starts to own the value at state $B_i$, i.e., $O(v_j,B_{i-1})=o_{t},O(v_j,B_i)=o_{t+1}$. Then, by the definition of proof, there exists a transaction in $b_i(o_t)$ that send the value to $o_{t+1}$. By the Rationality of Holding Value in RVO, $o_t$ would not make this transaction unless he would like to send this value. Then, by the Rationality of Sending Value in RVO, $o_t$ will take responsibility of giving proof $P(v_j,B_i)$ to $o_{t+1}$. Again, by the definition of proof, $P(v_j,B_i)$ is merely $P(v_j,B_k)\cup\{b_l(o_t): k<l \leq i\}\cup\{\mbox{public key of }o_t \}$, which can be independently provided by $o_t$. Hence, we prove that in this case $o_{t+1}$ will eventually has the proof $P(v_j,B_i)$. Furthermore, it is clear that only the owner of a value could transfer it as a transaction must be included in a block confirmed with the private key of the owner.

Then, we prove Liquidity. To transact a value, the owner simply needs to put a transaction in a confirmed transaction block. Then the property (Partial) Synchronous Liveness in Property~\ref{pro:bft} guarantees that the transaction block can be confirmed as the abstract will be included in the main chain.

At last, we prove Authenticity. This is actually guaranteed by the design of VAPOR. Firstly, the initial ownership of a value is unambiguous because it is on the main chain which has Asynchronous Consistency in Property~\ref{pro:bft}. Then, the ownership transition is always determined by a confirmed block which is immutable. Then, there are three possibilities for the number of transactions of the same value in a confirmed block: 1) if there is no transactions of that value, then the ownership remains unchanged; 2) if there is one transaction of that value, then the ownership is changed to the receiver; 3) if there are more than one transactions of that value, then the ownership becomes $NA$. Since all three possibilities result in unambiguous ownership, we proved Authenticity. \hfill \QEDA
\end{proof}

\section{Verification Algorithm for Value Division ${\tt GetOwnerDV}$}\label{a:godv}
Here we introduce ${\tt GetOwnerDV}$ in Algorithm~\ref{alg:godv}. Note that in here, a minor modification should be made on {\tt GetOwner} so that the result will not be `Fail' if redundant elements are detected in $p$.
\begin{algorithm}
\caption{Verification Algorithm for Divided Value ${\tt GetOwnerDV}(v_{[{\tt seq}]},B_i,p)$}\label{alg:godv}
\label{alg:tf}
\begin{algorithmic}
\State Find all value division transactions and their corresponding states in $p$. Order the states by $[s_1,s_2,\ldots]$;
\State $j \gets$ the first entry of $[{\tt seq}]$;
\State $t\gets 1$;
\While{$t \leq$ the length of {\tt seq}.} 
    \State ${\tt owner}={\tt GetOwner}(v_j, B_{s_1},p)$;
    \State Check if the corresponding value division transaction is in $b_{s_t}({\tt owner})$ and the sum of the amount of the divided value equals to the amount of the source value. Return `Fail' if the check fails. 
    \State $t++, j=[j,\mbox{next element in {\tt seq}}]$;
\EndWhile
\If{All blocks in $p$ are checked}
    \Return {\tt owner}
\Else
    \mbox{ }\Return Fail
\EndIf
\end{algorithmic}
\end{algorithm}

\section{Off-chain Payment Scheme}\label{a:offchain}

Our fast payment scheme contains two new type of transactions, two new types of message to the main chain, and a new verification algorithm ${\tt GetOwnerFP}$. If node $x$ wants to make fast payment to node $y$, he simply performs the following:
\begin{itemize}
    \item Node $x$ makes deposit transactions to lock up a number of values with indications that they could only be send to $y$, confirm the blocks, and send them to node $y$ to initialize the fast payment.
    \item When a fast payment of value $v_j$ is issued, node $x$ sends a signed transaction of $v_j$ to node $y$, denoted by $tx$. Then, node $y$ can include this transaction in his own blocks at any time and confirm them to receive the value.
    \item When node $x$ wants to end the fast payment and unlock a value $v_k$, he sends an unlock message to the main chain.
    \item The unlock will succeed in $T$ rounds if no objection message shows in the main chain. An objection message can be made by any node by sending $tx$ to the main chain.
\end{itemize}
Then, in ${\tt GetOwnerFP}$ we define three new rules on checking the proofs for ownership:
\begin{enumerate}
    \item A value $v_j$ locked by node $x$ is no longer considered as owned by $x$, but $NA$ indicating no owner. It will be reconsidered as owned by $x$ if there is only one unlock message is on the main chain, assume that it is included in $B_i$, and there is no objection message included in $B_k, i+1 \leq k \leq i+T$.
    \item A value $v_j$ is transacted from node $x$ to node $y$ in state $B_i$ if it is locked by node $x$ to send to node $y$ at a state $B_{i'}, i'<i$, and there is a signed transaction by $x$ included in block $b_{i}(y)$. There should not be a unlocking message for this value on the main chain that is not responded for more than $T$ blocks.
\end{enumerate}
Note that although a fast transaction is only confirmed when the block is confirmed, the transaction itself is completed as soon as the signed transaction is received by node $y$, since node $y$ can then independently make the proof of him owning this value. 

Some drawbacks in existing off-chain payment schemes, e.g., LN, are: 1), the values in the transactions and deposit will be locked until the channel is closed. Hence, it is a different type of transaction and can only be considered as a supplement to the value transfer system. 2), the receiver should have a certain synchrony, i.e., the receiver should be able to issue a transaction to the chain to take the deposit before it is refunded to the sender when he catches the sender cheating. 3), the security of this scheme is not formally proved.
A big advantage of the off-chain payment scheme in VAPOR is that node $y$ can spend $v_j$ as soon as he owns it, without requiring shutting down the whole channel, i.e., all deposit values been spend or unlocked. Moreover, we could use similar arguments as the proof in Subsection~\ref{ss:valvapor} to prove the Ownership property holds when the network is synchronous and the RVO rules apply.


\section{Betting Game}\label{a:betting}
Here, we give a smart contract for on-chain betting.
Node $x$ and node $y$ would like to bet even or odd on the hash of block $B_i$. Then, we simply add a new type of transaction which is $Bet: [v_j,x,y,B_i,sn]$. The bet transaction will lock the value $v_j$ until $B_i$ with one unlocking condition: another value with the same amount is bet by $y$ before $B_i$ with $x$ and the ownership will depend on the hash of $B_i$. Then, the verification algorithm is simply checking the lock transaction, the ownership for both values, and the hash of $B_i$, i.e., if node $x$ bet on even, then the ownership of both locked values will be node $x$ at state $B_i$ if the hash of $B_i$ is even.

However, the difficulty is to make sure that both node $x$ and node $y$ could get the proofs of ownership and the locking message for both values. This is a problem since there is always one node in the betting would benefit from not sharing the proof and/or the locking message, which will cause a scenario similar to Two Generals Problem. As a result, the verification algorithm must also check for a confirmation send by one node on the main chain, which shows the agreement for both nodes that both proofs are acquired. Without such confirmation, the value will be unlocked at state $B_i$ to its original owner.





\end{document}